\documentclass[a4paper,UKenglish,cleveref, autoref, thm-restate]{lipics-v2021}

\pdfoutput=1 %
\hideLIPIcs  %
\nolinenumbers

\usepackage{amsmath}
\usepackage{amsfonts}
\usepackage{hyperref}
\usepackage{graphicx}

\usepackage{tikz}
\usetikzlibrary{automata}
\usetikzlibrary{positioning}

\usepackage[colorinlistoftodos,color=red!10,prependcaption,textsize=small,
  disable %
]{todonotes}

\newcommand{\NN}{\mathbb{N}}

\newcommand{\SN}{\textsf{SN}}

\newcommand{\torel}{\mathrel{\to^=}}
\newcommand{\eps}{\epsilon}
\newcommand{\minf}{-}

\newcommand{\reminder}[2][1=]{}

\nolinenumbers %

\title{Old and New Benchmarks for Relative Termination of String Rewrite Systems}
\author{Dieter Hofbauer}{ASW Saarland}{message@dieter-hofbauer.de}{https://orcid.org/0000-0003-2094-6074}{}
\author{Johannes Waldmann}{HTWK Leipzig}{johannes.waldmann@htwk-leipzig.de}{}{}
\authorrunning{D.~Hofbauer and J.~Waldmann}


\begin{document}
\maketitle

\begin{abstract}
  We provide a critical assessment of the current set of benchmarks
  for relative SRS termination in the Termination Problems Database (TPDB):
  most of the benchmarks in \texttt{Waldmann\_19} and \texttt{ICFP\_10\_relative}
  are, in fact, strictly terminating
  (i.~e., terminating when non-strict rules are considered strict),
  so these benchmarks should be removed, or relabelled.
  To fill this gap, 
  we enumerate small relative string rewrite systems.
  At present, we have complete enumerations
  for a 2-letter alphabet up to size 11, and 
  for a 3-letter alphabet up to size 8.
  For some selected benchmarks, old and new, we discuss
  how to prove termination, automated or not. 

\end{abstract}

\section{Introduction}

A rewrite system $R$ is \emph{terminating relative to} a system $S$
if any (infinite) rewrite sequence modulo $R \cup S$ contains only
finitely many $R$-steps. This property is denoted by $\SN(R/S)$. 
The term was coined by Klop~\cite{Klop87}. 
Relative termination as a generalization of standard termination
is a basis for modular termination proofs
due to the fact that $\SN(R \cup S)$ holds if and only if
both $\SN(R/S)$ and $\SN(S)$ hold. 
It is a research topic on its own, 
cf.~\cite{DBLP:phd/dnb/Geser90, Zantema04}, 
and is widely used by current termination provers.

The Termination Problems Database (TPDB, 
see~\url{https://termination-portal.org/wiki/TPDB})
contains relative termination benchmarks since version~2.0.
For string rewriting, the directory \texttt{SRS\_Relative}
of the current TPDB (versions 11.*)
consists of five subdirectories: 

\begin{itemize}
\item \texttt{ICFP\_2010\_relative}:
  160 benchmarks, cf.~\cite{WaldmannFelgenhauer10} [VBS TC22: 132].

\item \texttt{Mixed\_relative\_SRS}:
  20 benchmarks %
  [VBS TC22: 20]. 
  
\item \texttt{Waldmann\_06\_relative}:
  11 benchmarks %
  [VBS TC22: 11]. 
  
\item \texttt{Waldmann\_19}:
  100 benchmarks,
  length-preserving (alphabet size 3, left- and right-hand sides of length 3,
  2 to 7 rules), randomly chosen
  [VBS TC22: 91]. 
  
\item \texttt{Zantema\_06\_relative}:
  14 benchmarks, hand-crafted %
  [VBS TC22: 9]. 
\end{itemize}

The number of benchmarks solved by the \emph{Virtual Best Solver (VBS)}
in the Termination Competition 2022 (TC22) are given in square brackets, 
see \url{https://termcomp.github.io/Y2022/SRS_Relative.html}.
``The VBS records the best (consistent) score for each claim
collected at least since 2018'', according to 
\url{https://termination-portal.org/wiki/Termination_Competition_2022}.

The category \texttt{SRS\_Relative} is missing an enumeration
of small systems.
We are starting this, and report some results and observations.
The first enumeration of (strict) SRS was given by Kurth~\cite{Kurth90},
and was later extended by Geser~\cite{Geser01}
and others~\cite{GeserWaldmannWenzel16}. 
Our ultimate goal is to cover all small relative SRSs (up to a certain size)
for which termination cannot be decided by current provers.
The naive approach is to enumerate them all,
and call the provers for each one.
The challenge is to achieve the same result with fewer prover calls,
by short-cutting the enumeration and omitting redundant cases.

There is no new theorem or method in this paper. 
It's just that no one has done the work before.
Source code and data for the experiments reported here are available at 
\url{https://gitlab.imn.htwk-leipzig.de/waldmann/pure-matchbox/-/issues/472}.

\section{How to %
  win \texttt{SRS\_Relative}, without proving Relative Termination}

The approach is strikingly simple: 
prove (standard) termination instead
for the \emph{strictified} system,
where all non-strict rules are replaced by their
strict counterparts
(replacing $\ell \torel r$ by $\ell \to r$). 
This is based on the trivial observation
that $\SN(R\cup S)$ implies $\SN(R/S)$.
Surprisingly, this works for at least 210
(out of 305) benchmarks in \texttt{SRS\_Relative}, 
in comparison to the winner of TC22, MultumNonMulta (MnM),
with 203 \texttt{YES} and 8 \texttt{NO} answers. 

What if the answer to the question about termination
of the strictified system $R\cup S$ is \texttt{NO}?
In this case we can infer $\neg\SN(R/S)$, 
provided we have a proof of $\SN(S)$.
This is in fact the method we recommend for proving $\SN(R/S)$:
if $\SN(S)$, then prove $\SN(R\cup S)$
using any of the methods for standard termination
that are not available for relative termination
(a DP transformation followed by weakly monotone matrix interpretations,
or RFC-matchbounds).
Only if $\neg\SN(S)$, check $\SN(R/S)$
with special methods (e.~g., strictly monotone matrix interpretations).

The point is that methods for standard termination are more powerful
(due to reduced monotonicity requirements, for example),
and that checking $\SN(S)$ is straightforward
(at least for the set of current benchmarks).
A potential drawback is that we cannot re-use
information from the proof, or disproof, of $\SN(S)$,
to investigate $\SN(R \cup S)$ or $\SN(R/S)$.

\section{Relative Non-Termination via Loops}

One way of proving the statement $\neg\SN(R/S)$ is 
to exhibit a mixed looping derivation
$v \to_{R\cup S}^+ u v w$
that contains at least one strict $R$-step.
In fact, all known non-terminating problems in \texttt{SRS\_Relative}
have been shown to be looping.

As shown by Geser and Zantema~\cite{DBLP:journals/ita/GeserZantema99},
loops for standard SRS can be characterized by looping forward closures. 
However, the corresponding statement does not hold for relative SRS: 
Consider the example
$\{a b \to a , c \torel b c \}$ from~\cite{DBLP:conf/rta/GeserHW19}, 
which has a loop ($abc \to ac \torel abc$),
but no looping forward closure.
In this example, a looping forward closure can be found
after reversing the system, but for the system
$\{b a b \to a , c \torel c b, d \torel b d \}$ 
neither the system nor its reversal has a looping forward closure,
although a loop exists ($c a d \torel^2 c b a b d \to c a d$). 

We could use overlap closures instead,
but this changes the nature of the enumeration
since there is one case where we need to overlap two closures with a rule.
If we want to limit ourselves to overlaps between one closure
and one rule, then we can
consider backward
overlaps.
A characterizing theorem for looping relative SRS is missing,
to the best of our knowledge. 

A special method for proving $\neg \SN(R/S)$ is this:
exhibit a derivation $v \to_S^+ u v w$
such that $u$ or $w$ contains an $R$-redex.
As an example consider $R = \{ a \to b \}$ and $S = \{ c \torel a c \}$,
where $c \to_S^+ a c$ and $a$ is an $R$-redex.
This corresponds to the so-called \emph{$R$-emitting loops}
for term rewriting, used by AProVE twice in TC22. 

\section{Experiments and  Results}

We did an optimized enumeration 
and called the prover matchbox on each relative SRS that was produced,
using a simple strategy that contains
  weights (from GLPK), 
  matrix interpretations of dimension 2 to 9 (with binary
  bitblasting for bit widths from 4 down to~1), 
  full tiling for width 2, 3, 4 (possibly repeated), and
  closure enumeration, 
using parallel computation on up to 8 cores.
Some unsolved problems will be submitted  to TPDB.

\subsection{Results for 2-letter alphabets}

All SRS up to size 9 could be solved.
Here, the size of a system is the sum of the sizes of its rules,
and the size of a rule is the sum of the lengths of its
left- and right hand sides. The empty string is denoted by $\eps$. 
Unsolved for size 10 are

\begin{minipage}{50mm}
  \begin{itemize}
  \item
    $\{ a a \to \eps , a b b \torel a b b b a \}$
  \end{itemize}
\end{minipage}
\qquad\qquad
\begin{minipage}{50mm}
  \begin{itemize}
  \item
    $\{ a a b b a \to b a a b, \eps \torel a \}$
  \end{itemize}
\end{minipage}

\noindent

\subsection{Results for 3-letter alphabets}

All SRS up to size 6 could be solved. 
Unsolved for size 7: 

\begin{minipage}{50mm}
  \begin{itemize}
  \item
    $\{ a c \to c, \eps \torel a b, a b \torel \eps \}$
  \end{itemize}
\end{minipage}
\qquad\qquad
\begin{minipage}{50mm}
  \begin{itemize}
  \item
    $\{ a c \to c, \eps \torel a b, b a \torel \eps \}$   
  \end{itemize}
\end{minipage}

\noindent
Unsolved for size 8: 24 systems, 
some of them looking vaguely similar, e.~g., 

\begin{minipage}{55mm}
  \begin{itemize}
  \item
    $\{ a c \to c, \eps \torel a b, a a b \torel \eps \}$, 
  \end{itemize}
\end{minipage}

\noindent
but also some that don't, like

\begin{minipage}{55mm}
  \begin{itemize}
  \item
    $\{ a b \to \eps , \eps \torel a c b, a c b \torel \eps \}$. 
  \end{itemize}
\end{minipage}

\section{Two Examples  from the Enumeration}

In this section, we give hand-crafted termination proofs
for two samples from the enumeration. 

\subsection{Example $\{ ac \to c , \eps \torel ab , ab \torel \eps  \}$} %

The following interpretation $M$ is a model for this system:
\begin{eqnarray*}
  \eps_M&=&0 \\
  a_M(y)&=&\max(0, y-1) \\
  b_M(y)&=&y+1 \\
  c_M(y)&=&0
\end{eqnarray*}
Here, a string $w\in\{a,b\}^*$ gets value 0 iff $w\to^*a^n$
via the relative rules alone; 
the number $n$ in that normal form gives the number of rewrite steps
starting from $wc$.
We compute $n$ as the first component of the monotone interpretation $I$
on domain $\NN^2$
(the second component being $M$),
ordered by $(x_1,y_1)>(x_2,y_2)$ iff $x_1>x_2\wedge y_1=y_2$: 
\begin{eqnarray*}
  \eps_I &=& (0,0) \\
  a_I(x,y) &=& \textsf{if} ~ y>0 ~ \textsf{then} ~ (x,y-1)  ~ \textsf{else} ~ (x+1,0) \\
  b_I(x,y) &=& (x,y+1) \\
  c_I(x,y) &=& (x,0)
\end{eqnarray*}

\noindent
Proof of Monotonicity, for $(x_1,y)>(x_2,y)$: 
\begin{eqnarray*}
  a_I(x_1,y) &=& \textsf{if} ~ y>0 ~ \textsf{then} ~ (x_1,y-1)  ~ \textsf{else} ~ (x_1+1,0) \\
             &>& \textsf{if} ~ y>0 ~ \textsf{then} ~ (x_2,y-1)  ~ \textsf{else} ~ (x_2+1,0) \\
             &=& a_I(x_2,y) 
\end{eqnarray*}

\noindent
Proof of Compatibility, again for $(x_1,y)>(x_2,y)$: 
\begin{eqnarray*}
  ac_I(x,y) &=& a_I(x,0)=(x+1,0) > (x,0)=c_I(x,y) \\
  ab_I(x,y) &=& a_I(x,y+1) = (x,y)
\end{eqnarray*}

\subsection{Example $\{ ac \to c , \eps \torel ab , ba \torel \eps  \}$} %

We want to use the same interpretation as in previous example, but
its second component
is no longer a model, since
$ba_I(x,0) = b_I(x+1,0)=(x+1,1)\neq (x,0)=\eps_I(x,0)$. 
The proof succeeds by keeping the interpretation and using different orders: 
\begin{eqnarray*}
  (x_1,y_1)>(x_2,y_2)
  &\text{iff}& (x_1> x_2) \wedge (y_1\ge y_2) \wedge (x_1-y_1 > x_2-y_2)  \\
  (x_1,y_1)\ge(x_2,y_2)
  &\text{iff}& (x_1\ge x_2) \wedge (y_1\ge y_2) \wedge (x_1-y_1\ge x_2-y_2) 
\end{eqnarray*}

\section{Some SRS from \texttt{Zantema\_06\_relative}}

We assume that these 13 systems were hand-crafted
to be ``obviously terminating'', 
but outside of the range of (then) current methods of automation.
Four of these benchmarks remain unsolved in the termination competition so far: 
\texttt{rel03}, \texttt{rel07}, \texttt{rel11}, and  \texttt{rel12}.
Meanwhile, we have automated proofs for \texttt{rel11} and \texttt{rel12},
obtained by brute-forced %
matrix interpretations~\cite{HofbauerWaldmann06, KoprowskiWaldmann08}. 
The following remain open:
\begin{itemize}
\item 
$\{ ac \to cca, c \torel baab, baab \torel c \}$
(\texttt{rel03})
\item
$\{ ad \to db , a \to bbb , d \to \eps, a \to \eps, 
bc \to cdd , ac \to bbcd, bdb \torel ad , ad \torel bdb \}$
(\texttt{rel07}). 
\end{itemize}

\subsection{\texttt{rel11}:
  $\{ b p b \to b a p b , p \torel a p a , a p a a \torel p \}$}

The following arctic-below-zero matrix interpretation of dimension 4
proves termination. 
Here, $-$ stands for $-\infty$: 
\[ 
  a \mapsto \begin{pmatrix}
      0& \minf& \minf& \minf\\
      \minf& -1& \minf& \minf\\
      \minf& \minf& 1& \minf\\
      \minf& \minf& \minf& 0
    \end{pmatrix}
    \qquad
    b \mapsto \begin{pmatrix}
      0& 0& \minf& \minf\\
      1& 1& \minf& 1\\
      0& 1& \minf& 1\\
      \minf& 0& \minf& 0
    \end{pmatrix}
    \qquad
    p \mapsto \begin{pmatrix}
      0& \minf& \minf& 0\\
      \minf& \minf& 1& \minf\\
      \minf& \minf& \minf& \minf\\
      \minf& \minf& \minf& 0
    \end{pmatrix}
\]

\subsection{\texttt{rel12}:
  $\{ b p b \to a b a p b a , p \torel a p a , a p a \torel p \}$}

A termination proof can be obtained by a
natural matrix interpretation of dimension 5: 
\[
  a \mapsto \begin{pmatrix}
      1& 0& 0& 0& 0\\
      0& 0& 1& 0& 0\\
      0& 0& 0& 1& 0\\
      0& 1& 0& 0& 0\\
      0& 0& 0& 0& 1
    \end{pmatrix}
    \qquad
    b \mapsto \begin{pmatrix}
      1& 1& 0& 0& 0\\
      0& 1& 4& 0& 1\\
      0& 1& 0& 0& 0\\
      0& 2& 0& 0& 0\\
      0& 0& 0& 0& 1
    \end{pmatrix}
    \qquad
    p \mapsto \begin{pmatrix}
      1& 0& 0& 0& 0\\
      0& 1& 0& 0& 0\\
      0& 0& 0& 1& 0\\
      0& 0& 1& 0& 0\\
      0& 0& 0& 0& 1
    \end{pmatrix}
\]

\section{Conclusion}

With this contribution, we provide some small hard benchmarks
for relative termination of string rewriting,
to encourage the search for more powerful proof methods
and their automation.

\end{document}